\documentclass{elsart}
\usepackage[square,numbers,sort&compress]{natbib}
\usepackage{amssymb}
\usepackage{graphicx}
\usepackage{amsmath}
\usepackage{calc}

\begin{document}
\begin{frontmatter}
\title{The influence of $(n-n')$-mixing processes in
$He^{*}(n) + He(1s^2)$ collisions on $He^{*}(n)$ atoms' populations
in weakly ionized helium plasmas}

\author[IF]{A. A. Mihajlov}
\author[IF]{Lj. M. Ignjatovi\'c\thanksref{contact}}
\thanks[contact]{e-mail: ljuba@phy.bg.ac.yu}
\author[IF]{V. A. Sre\'ckovi\'c}
\author[SILVACO]{Z. Djuri\'c}

\address[IF]{Institute of Physics, P.O.Box 57, Pregrevica 118, 11080 Zemun,
Belgrade, Serbia}
\address[SILVACO]{Silvaco Data Systems, Compass Point, St Ives PE27 5JL,
UK}

\begin{abstract}

The results of semi-classical calculations of rate coefficients of
$(n-n')$-mixing processes due to collisions of Rydberg atoms
$He^{*}(n)$ with $He(1s^2)$ atoms are presented. It is assumed that
these processes are caused by the resonant energy exchange within
the electron component of $He^{*}(n)+He$ collision system. The
method is realized through the numerical simulation of the
$(n-n')$-mixing processes, and is applied for calculations of the
corresponding rate coefficients. The calculations are performed for
the principal quantum numbers $n,\mbox{} n'$ in ranges $4 \le n < n'
\le 10$, and the atom and electron temperatures, $T_a, T_e$, in
domains $5000K \le T_{a} \le T_{e} \le 20000K$. It is shown that the
$(n-n')$-mixing processes can significantly influence the
populations of Rydberg atoms in non-equilibrium weakly ionized
helium plasmas with ionization degree $\sim 10^{-4}$. Therefore,
these  processes have to be included in the appropriate models of
such plasmas.

\end{abstract}
\end{frontmatter}

\section{Introduction}
\label{sec:intro}

Several existing models of collisional-radiative recombination
\cite{bat62a,bat64,bib67,bib79,ver81,mal86,bib87,per90,dju01} showed
that in the weakly ionized plasmas, with ionization degree $\sim
10^{-4}$, the electron transfer from the lowest atomic states to the
continuum and in the opposite direction is dominantly caused by the
cascade mechanism. These processes of excitation/de-excitation are
very important for populations in the lower part of the Rydberg's
block of states. The reason for this is the distribution function of
the excited atom states' populations in  weakly ionized plasmas,
which shows a distinct minimum ("bottleneck") in the lower part of
Rydberg's block of states. For example, paper \cite{mih04} compared
LTE (local thermodynamical equilibrium) and non-LTE distribution
functions, corresponding to different layers of the solar
photosphere \cite{ver81,mal86}. It was found that both functions had
distinct minimum in the domain of principal quantum numbers $n = 5 -
6$. Such behavior is illustrated in Fig.~\ref{fig:n_minimum},
showing LTE distribution functions for hydrogen plasmas at the
temperatures $T = 5000K$ and $6000K$.

Similar behavior of the distribution functions for excited atomic
states in the case of weakly ionized helium plasmas, for the
temperatures $T = 10000K$ and $12000K$ typical for photospheres of
some helium rich white dwarfs \cite{koe80}, is illustrated in the
same figure. The experimental data for weakly ionized plasma of non
equilibrium helium arc \cite{ale69}, show that the distribution
function has a similar shape, even in the case when the electron
temperature is four times higher than atomic temperature. It can be
concluded that the kinetics of weakly ionized hydrogen and helium
plasmas strongly depends on excitation and deexcitation processes in
the region $4 \lesssim n \lesssim 10$, where the "bottleneck" of
distribution function occurs.

The models of collisional-radiative recombination, used to estimate
the influence of non-elastic processes which populate the lower part
of Rydberg's block of states, included the radiation processes and
the excitation/de-excitation processes caused by electron-atom
collisions. The latter was assumed to be dominant for plasmas with
the ionization degree $\sim 10^{-4}$. However, the distribution of
excited atom states' populations experimentally obtained in
\cite{ale69}, showed a significant influence of $(n-n')$-mixing
processes in $He^{*}(n)+He(1s^{2})$ collisions. It was found that
the distribution function was similar to Boltzmann with the
temperature about $11000K$, close to the mid-value between
$T_{a}\cong 4500K$ and $T_{e}\cong 18000K$. Consequently, the
mechanism of the resonant energy exchange within the electron
component of the atom-Rydberg atom collision systems ({\it resonant
mechanism}) has been introduced in \cite{smi71} to explain such a
behavior. In cited paper and in \cite{jan79,mih82a} the resonant
mechanism was tested only for $(n-n')$-mixing processes in
$H^{*}(n)+H(1s)$ collisions, and the results obtained had only a
qualitative character.

The resonant mechanism has been successfully applied to
chemi-ionization processes in alkali Rydberg's atoms collisions with
their parents' ground state \cite{dev78}, and further for alkali
atoms' collisions \cite{mih81b,wei81,wei89,klu90,
bez97c,bez01,rya05,bet05,mic05,ign05}. In the case of hydrogen and
helium atoms, the resonant mechanism was used to analyze the
chemi-ionization and their inverse chemi-recombination processes
\cite{jan80,mih92,mih96,mih97b,mih97,mih03a,mih03b}. It was shown
that in weakly ionized hydrogen and helium plasmas the
chemi-ionization/recombination processes caused by the resonant
mechanism can be dominant in populating the Rydberg's states. This
mechanism was also investigated in the case of $(n-n')$-mixing
processes in $H^{*}(n)+H$ collisions populating the states within
the Rydberg's block \cite{mih04} (see also \cite{jan79,mih82a}). In
\cite{mih04} the corresponding rate coefficients were determined as
functions of $n$ and $T_{a}$. It was shown that in hydrogen plasma
with the ionization degree $\sim 10^{-4}$, and $4 \le n \le 10$, the
efficiency of these processes is comparable or even higher than the
concurrent excitation/de-excitation processes caused by
electron-atom collisions. This was confirmed later for weakly
ionized hydrogen plasma of solar photosphere in \cite{mih05a}.

The efficiency of the resonant mechanism for
chemi-ionization/recombination processes in weakly ionized hydrogen
and helium plasmas \cite{mih96,mih97b} suggested that one should
expect similar situation in the case of $(n-n')$-mixing processes.
The semi-classical method based on this mechanism, developed in
\cite{mih04}, was used in this paper and applied for the rate
coefficients' calculations of excitation processes

\begin{equation}
He^{*}(n) + He \rightarrow \left\{
             \begin{array}{l}
             \displaystyle{He^{*}(n'=n+p) + He },\\
             \displaystyle{ He + He^{*}(n'=n+p)},
             \end{array}
     \right.
\label{eq:ex}
\end{equation}
and their inverse deexcitation processes
\begin{equation}
He^{*}(n) + He \rightarrow \left\{
             \begin{array}{l}
             \displaystyle{He^{*}(n'=n-p) + He },\\
             \displaystyle{ He + He^{*}(n'=n-p)},
             \end{array}
     \right.
\label{eq:deex}
\end{equation}
with $p \ge 1$. The processes are characterized by the excitation
and deexcitation rate coefficients $K_{n;n+p}(T_{a})$ and
$K_{n;n-p}(T_{a})$, respectively, where $T_{a}$ is the atom
temperature. They represent the average characteristics of
transitions from all individual states with a given $n$ to all
states with given $n+p$ or $n-p$, and reflect the processes'
influence on the plasma kinetics. The calculations are performed for
the parameters' domains $4 \le n \le 10$, $1 \le p \le 5$ and $4000K
\le T_{a} \le T_{e} \le 20000K$. The efficiency of processes
(\ref{eq:ex}) and (\ref{eq:deex}) is compared to the concurrent
electron-excited atom collisional excitation/de-excitation processes
\begin{equation}
He^{*}(n) + e \rightarrow He^{*}(n \pm p) + e', \label{eq:Hee}
\end{equation}
where $e$ and $e'$ denote free electron in the initial and final
state, for given $n$, $p$ and electron $T_{e}$.

Beside of processes (\ref{eq:Hee}) in the plasma occur the other
concurrent processes, including the radiative processes caused by
the interaction of excited atoms with the electromagnetic radiation.
However, just the processes (\ref{eq:Hee}) are taken as referent
ones since they are always included in the modeling of weakly
ionized plasma disregarding to other concurrent processes.
Consequently, in order to demonstrate the necessity of including of
the processes (\ref{eq:ex}) and (\ref{eq:deex}) in the models of
weakly ionized plasmas, it is enough to show that their efficiency
is comparable with the efficiency of processes (\ref{eq:Hee}).

\section{Theoretical remarks}

\begin{figure}[h!]
\includegraphics[width=\columnwidth, height=0.75\columnwidth]{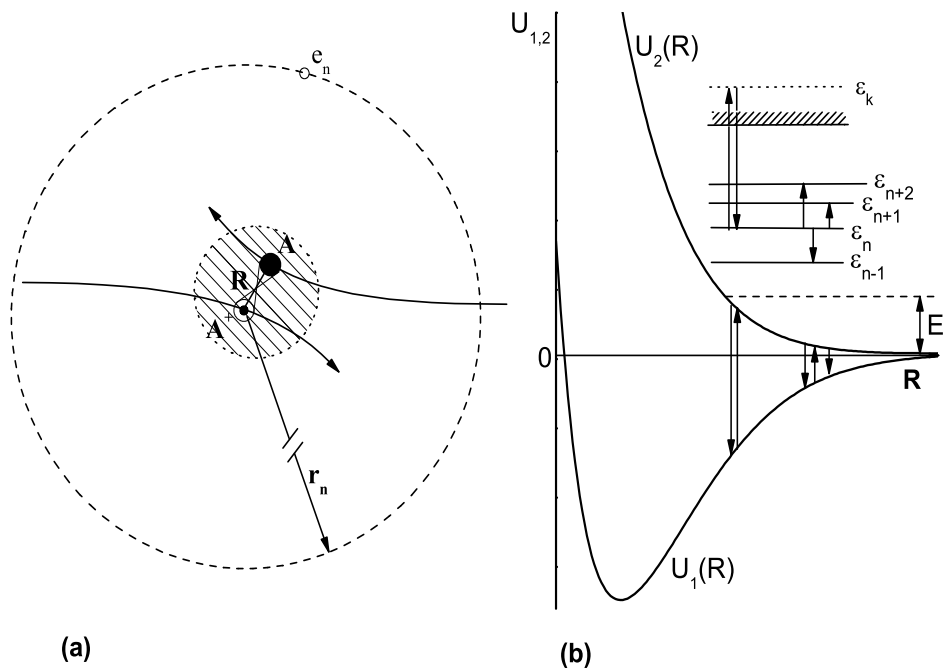}
\caption{\label{fig:mechanism} (a) Schematic illustration of
$He^{*}(n)+He$ collision (the region of the inter-nuclear distance
$R$ where the outer electron is collectivized is shaded); (b)
Schematic illustration of the simultaneous resonant transitions of
the outer electron from the initial bound to the final free state
and the sub-system $He^{+}+He$ from initial excited to the final
ground electronic state.}
\end{figure}

{\bf Resonant mechanism.} The resonant mechanism of non-elastic
$(n-n')$-mixing and chemi-ionization/recombination processes in
$A^{*}(n)+A$ collisions was described in details in previous papers
\cite{jan79,jan80,mih81b,mih92,mih04}. In considered cases $A$ was
an atom in the ground state with one or two $s$-electrons out of
completely filled shells. The mechanism, briefly described here, is
illustrated by Fig.~\ref{fig:mechanism}a and
Fig.~\ref{fig:mechanism}b.

Fig.~\ref{fig:mechanism}a shows that the resonant mechanism's region is
defined by the inequality
\begin{equation}
R \ll r_{n}, \label{eq:rn}
\end{equation}
where $R$ is the internuclear distance and $r_{n}\sim n^{2}$ the
characteristic radius of Rydberg's atom $A^{*}(n)$. In this region a
perturbation operator takes the dipole part of the interaction between the
outer electron $e_{n}$ and the sub-system $A^{+} + A$.

The electronic state of the $A^{+} + A$ sub-system is described by
the wave functions of the adiabatic ground ($|1;R\rangle$) and first
excited ($|2;R\rangle$) electronic states of the molecular ion
$A_{2}^{+}$. The potential curves of these states are denoted with
$U_{1}(R)$ and $U_{2}(R)$, and schematically shown in
Fig.~\ref{fig:mechanism}b. The impact energy of the collisional
system, determined in the center of mass' system, is denoted with
$E$. Fig.~\ref{fig:mechanism}b illustrates that the resonant
mechanism takes into account the transitions $|n\rangle \rightarrow
|n+p\rangle$, $|n\rangle \rightarrow |\vec{k}\rangle$ and
$|\vec{k}\rangle \rightarrow |n\rangle$, $|n\rangle \rightarrow
|n-p\rangle$ of the outer electron, which occur simultaneously with
the transitions $|2;R\rangle \rightarrow |1;R\rangle$ and
$|1;R\rangle \rightarrow |2;R\rangle$ in the sub-system $A^{+} + A$.

The main cause of the resonant character of described mechanism is the
importance of transitions in the close vicinity of the
corresponding resonant distances. In the case of $(n-n')$-mixing, the
resonant  distances $R_{n;n\pm p}$ are found as roots of the equation
\begin{equation}
U_{12}(R) \equiv U_{2}(R)-U_{1}(R)=|\epsilon_{n \pm
p}-\epsilon_{n}|,
\label{eq:koren}
\end{equation}
where $\epsilon_{n}$ and $\epsilon_{n \pm p}$ denote electron energies in
bound states with principal quantum numbers $n$ and $n \pm p$. Note that
for the chemi-ionization processes the resonant distances $R_{n;k}$ are
found from (\ref{eq:koren}) by replacing $\epsilon_{n \pm p}$ with the
energy of the outer electron in free state, $\epsilon_{k}$.

The described resonant mechanism for $(n-n')$-mixing processes
(\ref{eq:ex}) and (\ref{eq:deex}) is applied in helium case: $|1;R\rangle$
and $|2;R\rangle$ are $X^{2}\Sigma_{u}^{+}$- and
$A^{2}\Sigma_{g}^{+}$-states of the molecular ion $He_{2}^{+}$. The
adiabatic potential curves $U_{1}(R)$ and $U_{2}(R)$ are taken from
\cite{gup67,met92}. The  resonant distances $R_{n;n\pm p}$ are determined
from Eq.~(\ref{eq:koren}) with these potential curves. The values of
$R_{n;n + p}$ for $4 \le n \le 10$ and $1 \le p \le 5$ are presented in
Table~\ref{tab:Rnp}.

The rate coefficients' calculations require the
quantity $D_{12}(R)=|\langle1;R|{\bf D}|2;R\rangle|$, where ${\bf D}$ is
the operator of the dipole momentum of the molecular ion $He_{2}^{+}$.
Table~\ref{tab:Rnp} shows that for  all values of $R_{n;n + p}$ the
inequality $R \gg a_{0}$ holds ($a_{0}$ being the atomic unit of length),
and for $D_{12}$ the approximation from \cite{mih04} can be used
\begin{equation}
D_{12}(R) = \frac{e \cdot R}{2}, \label{eq:D12}
\end{equation}
where $e$ is the absolute value of electron charge.

{\bf The method of calculations.} $He^{*}(n \ge 4) + He$ and
$H^{*}(n \ge 4) + H$ collision systems have similar behavior in $R
\gg a_{0}$ region, which has been used for the chemi-ionization
processes' calculations \cite{mih97b,mih03b}. On the other hand, the
resonant mechanism generates transitions between Rydberg's states of
outer electron that are allowed by selective rules. This was used to
apply a semi-classical theory for the chemi-ionization processes in
the cases of hydrogen \cite{mih92,mih96} and helium
\cite{mih97b,mih03b}.

The $(n-n')$-mixing processes in $He^{*}(n)+He$ collisions are
considered to be a result of the same resonant mechanism as in
previous cases. The similarities between hydrogen and helium
collisional systems made it possible to apply a semi-classical
theory from \cite{mih04} for processes (\ref{eq:ex}) and
(\ref{eq:deex}). It means that the rate coefficients for these
processes can be determined by the same procedure used in
\cite{mih04}, but with the molecular ion $He_{2}^{+}$
characteristics instead. Consequently, the rate coefficients
$K_{n;n+p}(T_{a})$ for excitation processes (\ref{eq:ex}) are
determined directly, and the coefficients $K_{n;n-p}(T_{a})$ for
deexcitation processes (\ref{eq:deex}) by using thermodynamical
balance principle.

Similar to \cite{mih04}, the rate coefficients $K_{n;n+p}(T_{a})$ are
calculated under assumptions that:
\begin{itemize}
    \item the internuclear motion occurs with the probability $1/2$ in the
    potential $U_{2}(R)$ (molecular state $|2;R\rangle$), and can be
    described by the classical trajectory determined by the impact
    parameter and energy $E$;

    \item the transitions $|2;R\rangle|n\rangle \rightarrow
    |1;R\rangle|n+p\rangle$ for $p=1,2...$ happen continually along the
    trajectory, and are characterized by their rates;

    \item the atomic component has a Maxwell distribution described by the impact energies
    $E$ with a given temperature $T_{a}$.
\end{itemize}

Applying the method from \cite{mih04}, the excitation rate coefficients
$K_{n;n+p}(T_{a})$ are given as
\begin{equation}
\begin{array}{l}
{\displaystyle K_{n;n+p}(T_{a})=\frac{2
\pi}{3\sqrt{3}}\frac{(ea_{0})^{2}}{\hbar}
\cdot n^{-5} \cdot g_{n;n+p} \times } \\
{\displaystyle \times \int\limits_{R_{min}(n,n+p)}^{R_{max}(n,n+p)}
X(R)\cdot\exp\left[-\frac{U_{2}(R)}{kT_{a}}\right]\frac{R^{4}\cdot
dR}{a_{0}^{5}},}
\end{array}
\label{eq:Knn}
\end{equation}
where $g_{n;n+p}$ is Gaunt's factor for $|n\rangle \rightarrow
|n+p\rangle$ transition, determined in \cite{joh72}. The lower and
upper bounds $R_{min}(n,n+p)$ and $R_{max}(n,n+p)$ are found as roots
of the equations
\begin{equation}
U_{12}(R=R_{min})=\epsilon_{n+p+1-\Delta p}-\epsilon_{n}, \qquad
U_{12}(R=R_{max})=\epsilon_{n+p-\Delta p}-\epsilon_{n}
\label{eq:Rmin}
\end{equation}
with $\epsilon_{n}=-Ry/n^{2}$ and $\Delta p=0.380$. A factor $X(R)$
is given by
\begin{equation}
X(R)=\frac{\Gamma (\frac{3}{2},\frac{-U_{1}(R)}{kT_{a}})}{\Gamma
(\frac{3}{2})},
\label{eq:X}
\end{equation}
and $\Gamma (x,y)$ and $\Gamma (x)$ are incomplete and complete
gamma functions \cite{abr65}.

The deexcitation rate coefficients $K_{n-p;n}(T_{a})$ are calculated
from the thermodynamic balance equation in the case of helium plasma
in the state of thermodynamical equilibrium with $T=T_{a}$, namely
\begin{equation}K_{n-p;n}(T=T_{a})N_{eq}(n-p)N_{eq}(1)=
K_{n;n-p}(T=T_{a})N_{eq}(n)N_{eq}(1),
\label{eq:bal}
\end{equation}
where $N_{eq}(1)$, $N_{eq}(n-p)$ and $N_{eq}(n)$ are the
corresponding equilibrium densities of the atoms $He(1s^{2})$,
$He^{*}(n-p)$ and $He^{*}(n)$, respectively. These densities satisfy
equations
\begin{equation}
\begin{array}{l}
\displaystyle{ N_{eq}(n-p)=N_{eq}(1)\cdot 4(n-p)^{2}
\exp{\left(-\frac{I_{He}+\epsilon_{n-p}}{kT_{a}}\right)},} \\
\displaystyle{ N_{eq}(n)=N_{eq}(1)\cdot 4n^{2}
\exp{\left(-\frac{I_{He}+\epsilon_{n}}{kT_{a}}\right)},}
\end{array}
\label{eq:bal1}
\end{equation}
where $I_{He}$ is the ionization potential of $He(1s^{2})$ atom.
Consequently, the relation which links the excitation and
deexcitation rate coefficients is
\begin{equation}
K_{n;n-p}(T_{a})=K_{n-p;n}(T_{a})\cdot \frac{(n-p)^{2}}{n^{2}}\cdot
\exp{\left(\frac{\epsilon_{n-p;n}}{kT_{a}}\right)},
\label{eq:Kdex}
\end{equation}
where $\epsilon_{n-p;n}=\epsilon_{n}-\epsilon_{n-p}$. The usage of
the thermodynamical balance principle understands that in the
equilibrium helium plasma with $T=T_{a}$ Beoltzmann's distribution
of helium excited atom populations has its full physical sense. One
can directly confirms itself that this is valid for $T_{a} \gtrsim
3000K$. Because of that the described procedure can be certainly
applied in the temperature region considered here.

\section{Results and discussion}

Equations~(\ref{eq:Knn})-(\ref{eq:X}) are used for rate coefficients
$K_{n;n+p}(T_{a})$ calculations. The results for $4 \le n \le 10$, $1 \le p
\le 5$ and $4000K \le T_{a} \le 20000K$ are presented in
Tab.~\ref{tab:Knp}. The Table shows a monotonous decrease of
$K_{n;n+p}(T_{a})$ with the increases of $n$ and $p$, and a very slow
increase as $T_{a}$ increases. It should be noted that the differences
between $K_{n;n+p}$ values for hydrogen from \cite{mih04} and these for
helium are caused by differences between the resonant distances
$R_{n;n+p}$. The deexcitation rate coefficients $K_{n;n-p}(T_{a})$ for $4 <
n \le 10$ and $4 \le n-p$ are calculated according to Eq.~(\ref{eq:Kdex})
and using the values for excitation rate coefficient from
Tab.~\ref{tab:Knp}.

The relative efficiency of $(n-n')$-mixing processes (\ref{eq:ex}) and
(\ref{eq:deex}) in comparison with  electron-atom collision processes
(\ref{eq:Hee}) is characterized by the quantity $F_{n;n\pm p}(T_{a},T_{e})$
\begin{equation}
F_{n;n \pm p}(T_{a},T_{e})=\frac{K_{n;n \pm p}(T_{a})N(n)N(1)}
{\alpha_{n;n \pm p}(T_{e})N(n)N_{e}}=\frac{K_{n;n \pm
p}(T_{a})}{\alpha_{n;n \pm p}(T_{e})}\cdot \eta_{ea},
\label{eq:F}
\end{equation}
\begin{equation}
\eta_{ea}=\frac{N(1)}{N_{e}}, \label{eq:eta}
\end{equation}
where $\alpha_{n,n \pm p}(T_{e})$ are the rate coefficients for the
electron-atom excitation processes (\ref{eq:Hee}), and $N_{e}$ is
the free electron density. In the case of weakly ionized helium
plasma the parameter $\eta$ is closed to (ionization degree)$^{-1}$.
The values for $\alpha_{n,n \pm p}(T_{e})$ are obtained from
\cite{fuj78,jan87}.

\begin{figure}[h!]
\includegraphics[width=\columnwidth, height=0.75\columnwidth]{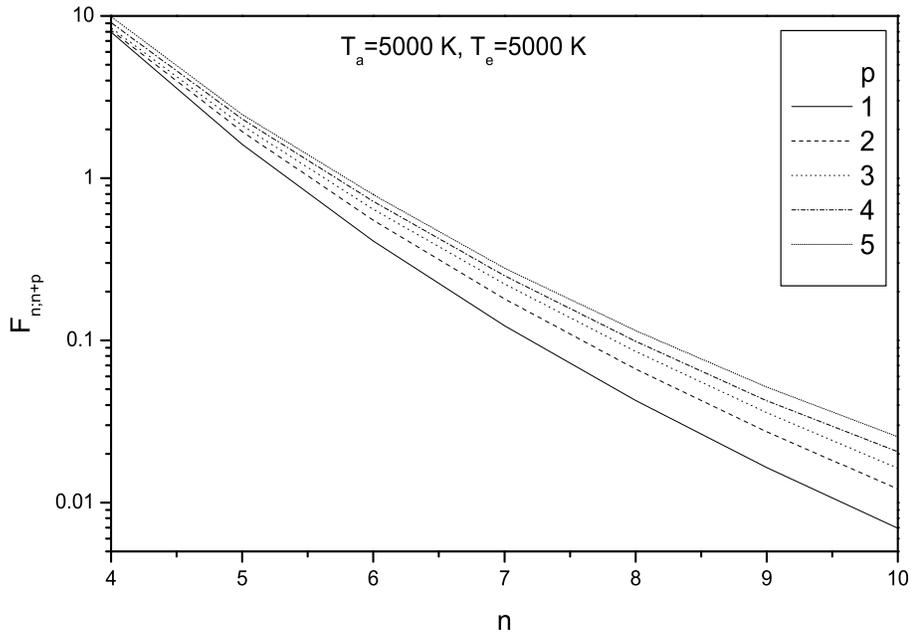}
\caption{\label{fig:F5_5} Parameter $F_{n,n+p}$ for $T_{a}=5000K$ and
$T_{e}=5000K$.}
\end{figure}

\begin{figure}[h!]
\includegraphics[width=\columnwidth, height=0.75\columnwidth]{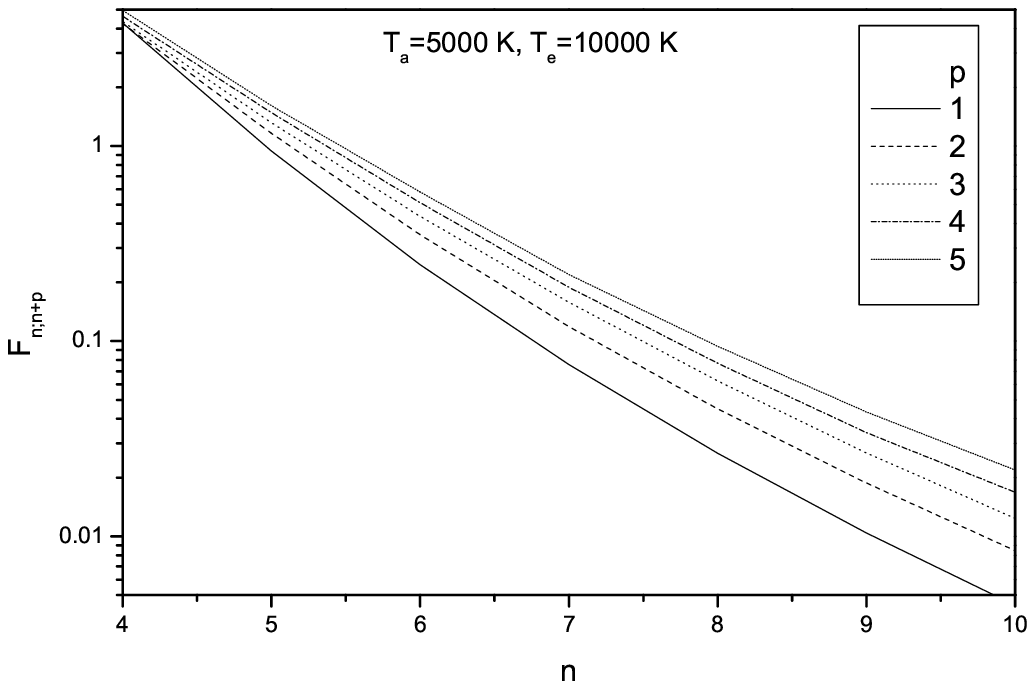}
\caption{\label{fig:F5_10} Same as in Fig.~\ref{fig:F5_5}, but for
$T_{a}=5000K$ and $T_{e}=10000K$.}
\end{figure}

\begin{figure}[h!]
\includegraphics[width=\columnwidth, height=0.75\columnwidth]{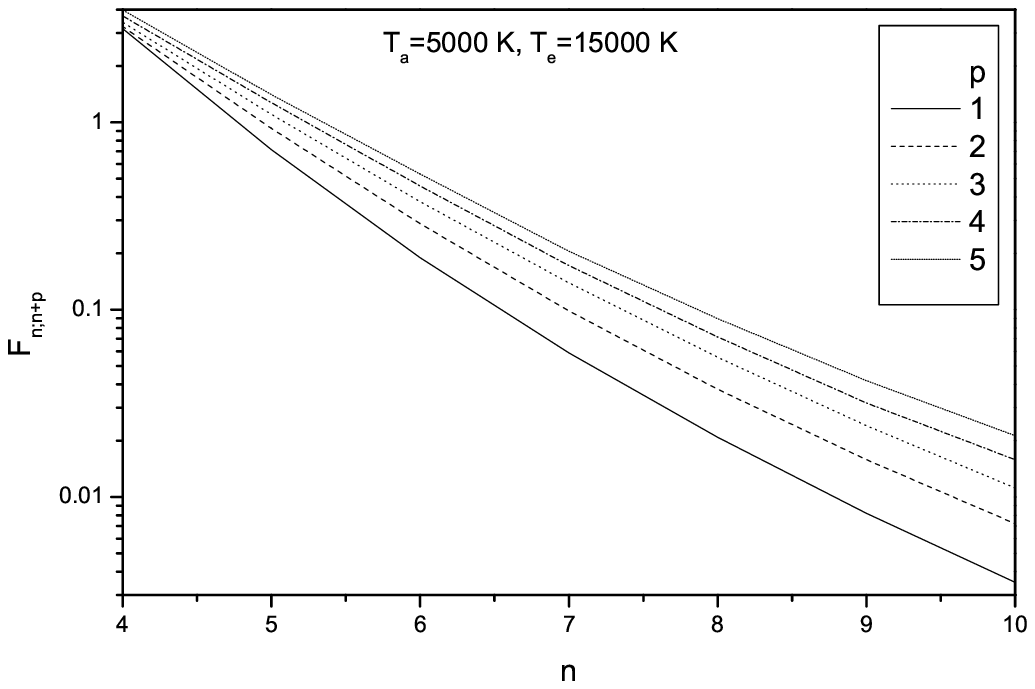}
\caption{\label{fig:F5_15} Same as in Fig.~\ref{fig:F5_5}, but for
$T_{a}=5000K$ and $T_{e}=15000K$.}
\end{figure}

\begin{figure}[h!]
\includegraphics[width=\columnwidth, height=0.75\columnwidth]{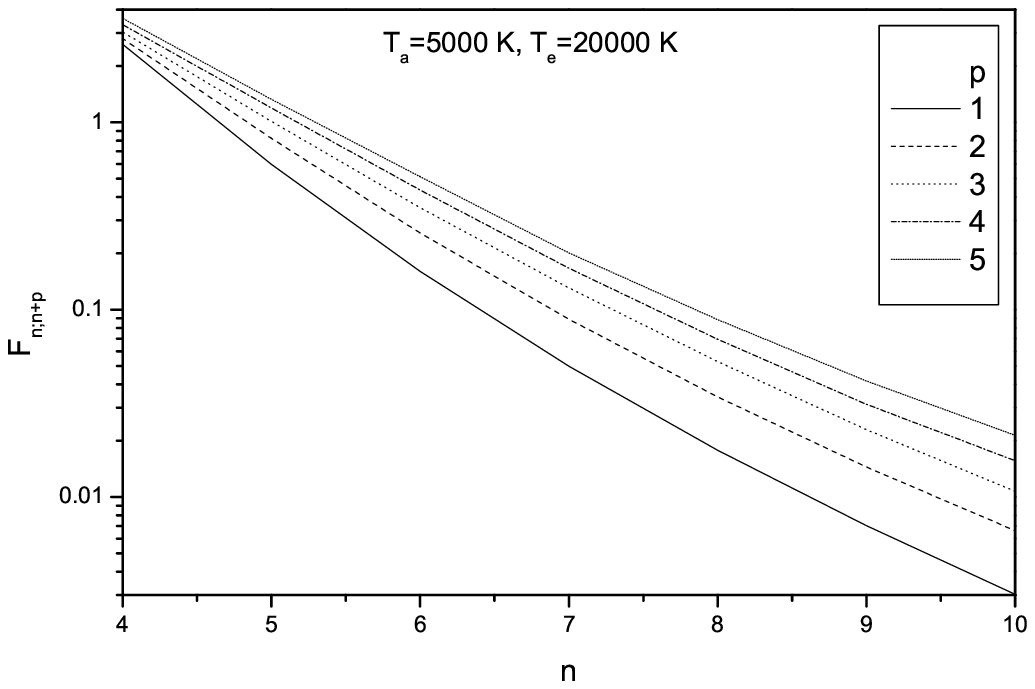}
\caption{\label{fig:F5_20}
same as in Fig.~\ref{fig:F5_5}, but for $T_{a}=5000K$ and $T_{e}=20000K$.}
\end{figure}

\begin{figure}[h!]
\includegraphics[width=\columnwidth, height=0.75\columnwidth]{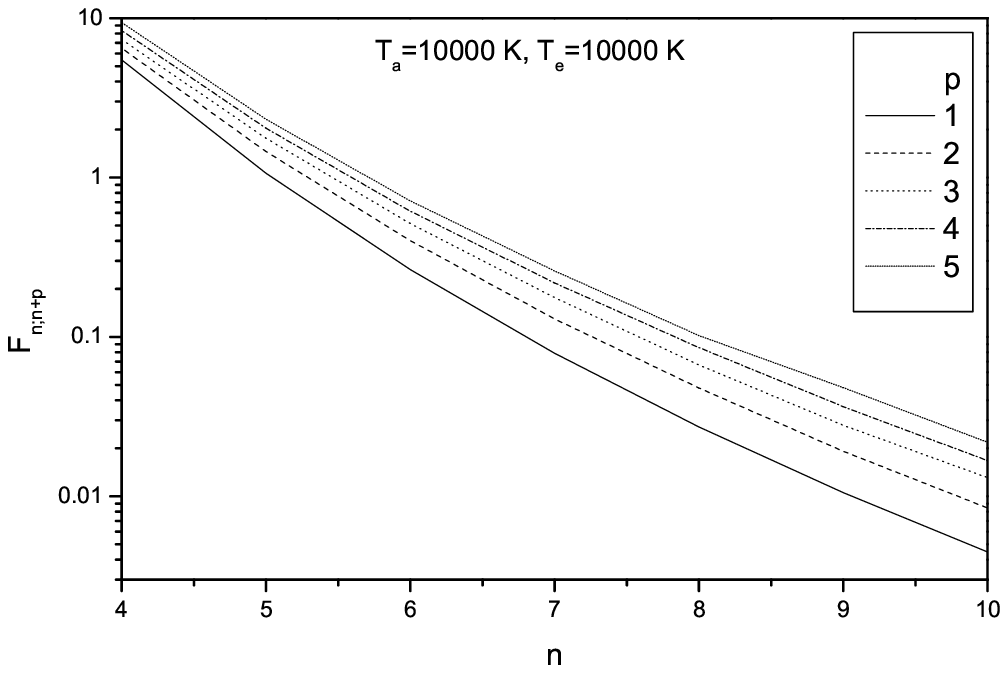}
\caption{\label{fig:F10_10} Same as in Fig.~\ref{fig:F5_5}, but for
$T_{a}=10000K$ and $T_{e}=10000K$.}
\end{figure}

\begin{figure}[h!]
\includegraphics[width=\columnwidth, height=0.75\columnwidth]{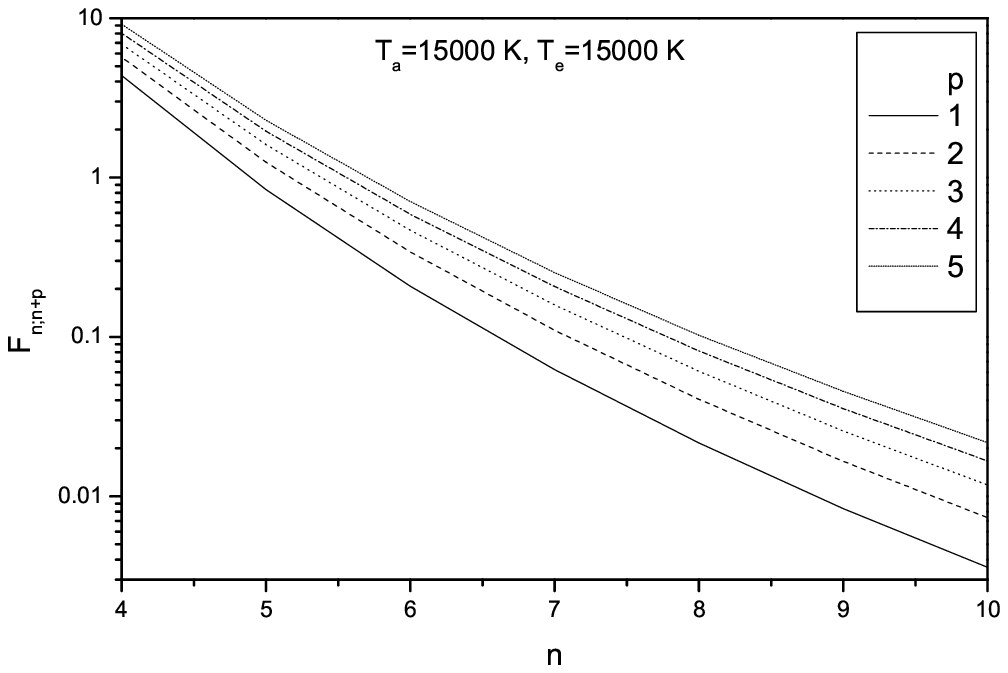}
\caption{\label{fig:F15_15} Same as in Fig.~\ref{fig:F5_5}, but for
$T_{a}=15000K$ and $T_{e}=15000K$.}
\end{figure}

\begin{figure}[h!]
\includegraphics[width=\columnwidth, height=0.75\columnwidth]{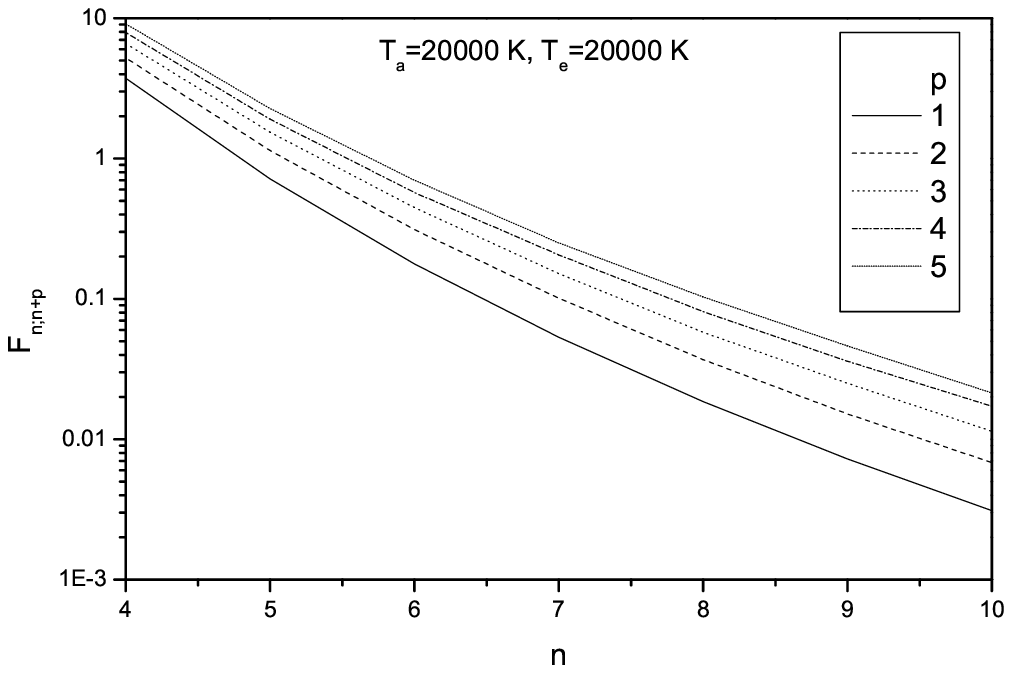}
\caption{\label{fig:F20_20} Same as in Fig.~\ref{fig:F5_5}, but for
$T_{a}=20000K$ and $T_{e}=20000K$.}
\end{figure}

The quantities $F_{n;n\pm p}(T_{a},T_{e})$ as functions of $n$ for given
$p$, $T_{a}$ and $T_{e}$ are illustrated in
Figs.~\ref{fig:F5_5}-\ref{fig:F20_20}. They show that the efficiency of
$(n-n')$-mixing processes in $He^{*}(n)+He$ collisions, in domains $4 \le n
\le 8$ and $n+1 \le n' \le n+5$, is higher or at least comparable to that
of electron-atom processes (\ref{eq:Hee}), for all considered $T_{a}$ and
$T_{e}$.

\begin{figure}[h!]
\includegraphics[width=\columnwidth, height=0.75\columnwidth]{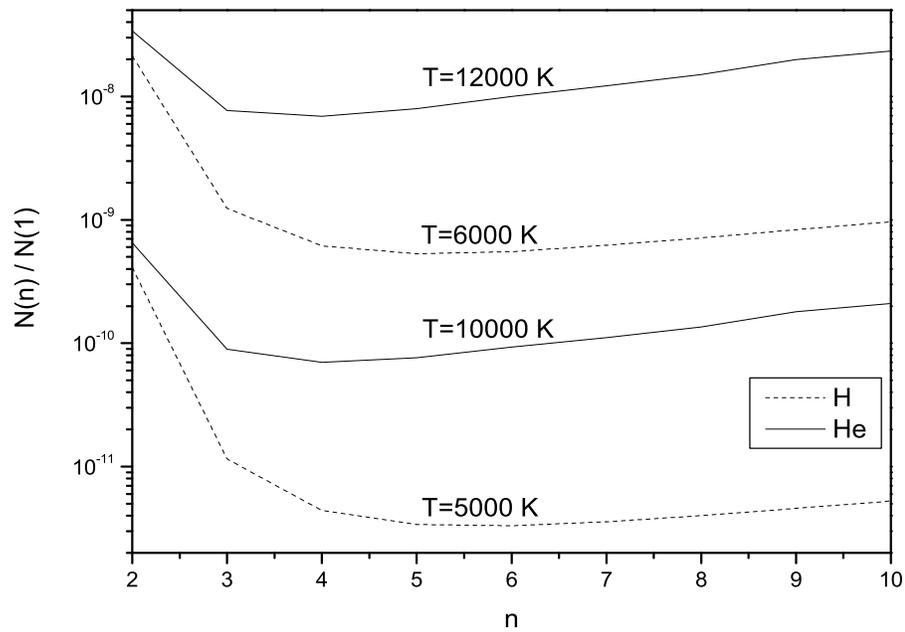}
\caption{\label{fig:n_minimum} The distribution function of excited
atom population.}
\end{figure}

The region $4 \le n \le 6$, where the minimum of helium excited atom
populations' distribution occurs, is particularly important (see
Fig.~\ref{fig:n_minimum}). In this domain the efficiency of
processes (\ref{eq:ex}) and (\ref{eq:deex}) is significantly higher
or very close to that of electron-atom processes (\ref{eq:Hee}).
Such efficiency can explain the shape of distribution function,
experimentally obtained in \cite{ale69}, and discussed in
Sec.~\ref{sec:intro}.

Regarding to the presented method accuracy we rely to the fact that
(n-n')-mixing processes are caused by the same resonant mechanism as
the above mentioned chemi-ionization processes in symmetric
atom-Rydberg atom collisions. Namely, these processes are well
studied experimentally and theoretically and it was found that there
is very good agreement between experimental and theoretical results
\cite{mih81b,wei81,wei89,klu90,
bez97c,bez01,rya05,bet05,mic05,ign05}.

The results of this paper suggest that the influence of the
processes (\ref{eq:ex}) and (\ref{eq:deex}) on populating the lower
part of Rydberg block of states has to be taken into account. It was
shown in \cite{mih03a} that the similar mechanism in the
chemi-ionization processes in weakly ionized hydrogen plasma, in
region $4 \le n \le 8$, significantly changed the excited atom
populations in the entire Rydberg block of states, and hence the
changes in atomic spectral line shapes \cite{mih07b}. Therefore, the
role of $(n-n')$-mixing processes for the populations of the lower
part of Rydberg block of states is of a special importance, the fact
that has already been confirmed in the case of hydrogen
\cite{mih04}.

The importance of chemi-ionization/recombination processes in weakly
ionized helium plasmas has been confirmed earlier
\cite{mih97b,mih03b}, particularly when they were included within a
model of collision radiative recombination \cite{dju01}. On the
other hands, the results presented here suggest a similar role for
$(n-n')$-mixing processes (\ref{eq:ex}) and (\ref{eq:deex}) in both
helium and hydrogen plasmas. For example, these processes should
influence helium spectral lines' shapes in cases of some helium rich
white dwarfs atmospheres, in a research which already started
\cite{bea97}. They are also important for the kinetics of weakly
ionized laboratory plasmas, which are always used in a research of
gaseous discharges and inner-plasma collision processes
\cite{ale69,ale74,pen83,kis87,bay89,sol93,den99}. To conclude, the
$(n-n')$-mixing processes (\ref{eq:ex}) and (\ref{eq:deex}) have to
be included in modeling of both laboratory and astrophysical weakly
ionized helium plasmas.


\newcommand{\noopsort}[1]{} \newcommand{\printfirst}[2]{#1}
  \newcommand{\singleletter}[1]{#1} \newcommand{\switchargs}[2]{#2#1}

\newpage
\linespread{0.75}
\begin{table}
\caption{$R_{n;n+p}/a_{0}$ values calculated from (\ref{eq:koren})
for $n'=n+p$.}
\begin{tabular}{@{} c c c c c c @{} } \hline\hline
\multicolumn{1}{c}{} & \multicolumn{5}{c}{$p$}
\\ \cline{2-6}
 $n$&   1   &   2   &   3   &   4   &   5   \\
\hline\hline
  4 & 4.907 & 4.556 & 4.400 & 4.312 & 4.257 \\
  5 & 5.398 & 5.019 & 4.843 & 4.741 & 4.675 \\
  6 & 5.802 & 5.403 & 5.211 & 5.097 & 5.022 \\
  7 & 6.146 & 5.731 & 5.527 & 5.403 & 5.321 \\
  8 & 6.446 & 6.018 & 5.803 & 5.672 & 5.582 \\
  9 & 6.711 & 6.272 & 6.050 & 5.911 & 5.816 \\
 10 & 6.949 & 6.502 & 6.272 & 6.128 & 6.028 \\
\hline\hline
\end{tabular}
\label{tab:Rnp}
\end{table}
\begin{table}
\caption{Excitation rate coefficients
$K_{n;n+p}(T_{a})$[$10^{-9}cm^{3}s^{-1}$].}
\begin{tabular}{@{} c c c c c c c c c c c @{} } \hline\hline
\multicolumn{1}{c}{} & \multicolumn{1}{c}{} &
\multicolumn{9}{c}{$T_{a}[10^{3}K]$}
\\ \cline{3-11}
 $n$& $p$&  4.0 &  4.5  &  5.0  &  6.0  &  8.0  & 10.0  &  12.0 &  16.0 &  20.0 \\
\hline\hline
    & 1 & 1.005 & 1.080 & 1.143 & 1.243 & 1.378 & 1.462 & 1.520 & 1.594 & 1.638 \\
    & 2 & 0.195 & 0.221 & 0.244 & 0.283 & 0.338 & 0.374 & 0.400 & 0.434 & 0.455 \\
  4 & 3 & 0.072 & 0.084 & 0.095 & 0.114 & 0.142 & 0.161 & 0.175 & 0.194 & 0.206 \\
    & 4 & 0.035 & 0.041 & 0.048 & 0.058 & 0.075 & 0.087 & 0.095 & 0.107 & 0.114 \\
    & 5 & 0.020 & 0.024 & 0.028 & 0.035 & 0.045 & 0.053 & 0.058 & 0.066 & 0.071 \\
\hline\hline
    & 1 & 0.682 & 0.707 & 0.728 & 0.760 & 0.800 & 0.824 & 0.841 & 0.861 & 0.873 \\
    & 2 & 0.170 & 0.182 & 0.191 & 0.207 & 0.227 & 0.239 & 0.248 & 0.258 & 0.265 \\
  5 & 3 & 0.072 & 0.078 & 0.083 & 0.092 & 0.103 & 0.111 & 0.116 & 0.122 & 0.126 \\
    & 4 & 0.038 & 0.042 & 0.045 & 0.050 & 0.058 & 0.062 & 0.066 & 0.070 & 0.072 \\
    & 5 & 0.023 & 0.025 & 0.027 & 0.031 & 0.036 & 0.039 & 0.041 & 0.044 & 0.046 \\
\hline\hline
    & 1 & 0.432 & 0.441 & 0.448 & 0.459 & 0.473 & 0.481 & 0.486 & 0.493 & 0.497 \\
    & 2 & 0.124 & 0.128 & 0.132 & 0.138 & 0.146 & 0.151 & 0.154 & 0.157 & 0.160 \\
  6 & 3 & 0.056 & 0.059 & 0.062 & 0.065 & 0.070 & 0.073 & 0.075 & 0.077 & 0.079 \\
    & 4 & 0.031 & 0.033 & 0.035 & 0.037 & 0.040 & 0.042 & 0.044 & 0.045 & 0.046 \\
    & 5 & 0.019 & 0.021 & 0.022 & 0.024 & 0.026 & 0.027 & 0.028 & 0.029 & 0.030 \\
\hline\hline
    & 1 & 0.276 & 0.279 & 0.282 & 0.286 & 0.292 & 0.295 & 0.297 & 0.299 & 0.301 \\
    & 2 & 0.086 & 0.088 & 0.089 & 0.092 & 0.095 & 0.097 & 0.098 & 0.100 & 0.101 \\
  7 & 3 & 0.041 & 0.042 & 0.044 & 0.045 & 0.047 & 0.049 & 0.049 & 0.050 & 0.051 \\
    & 4 & 0.024 & 0.025 & 0.025 & 0.027 & 0.028 & 0.029 & 0.029 & 0.030 & 0.031 \\
    & 5 & 0.015 & 0.016 & 0.016 & 0.017 & 0.018 & 0.019 & 0.019 & 0.020 & 0.020 \\
\hline\hline
    & 1 & 0.181 & 0.183 & 0.184 & 0.186 & 0.188 & 0.189 & 0.190 & 0.191 & 0.192 \\
    & 2 & 0.059 & 0.060 & 0.061 & 0.062 & 0.064 & 0.065 & 0.065 & 0.066 & 0.066 \\
  8 & 3 & 0.030 & 0.030 & 0.031 & 0.032 & 0.033 & 0.033 & 0.034 & 0.034 & 0.034 \\
    & 4 & 0.018 & 0.018 & 0.018 & 0.019 & 0.020 & 0.020 & 0.020 & 0.021 & 0.021 \\
    & 5 & 0.011 & 0.012 & 0.012 & 0.013 & 0.013 & 0.013 & 0.014 & 0.014 & 0.014 \\
\hline\hline
    & 1 & 0.122 & 0.123 & 0.124 & 0.125 & 0.126 & 0.126 & 0.127 & 0.127 & 0.128 \\
    & 2 & 0.042 & 0.042 & 0.043 & 0.043 & 0.044 & 0.044 & 0.045 & 0.045 & 0.045 \\
  9 & 3 & 0.021 & 0.022 & 0.022 & 0.022 & 0.023 & 0.023 & 0.023 & 0.024 & 0.024 \\
    & 4 & 0.013 & 0.013 & 0.013 & 0.014 & 0.014 & 0.014 & 0.014 & 0.015 & 0.015 \\
    & 5 & 0.009 & 0.009 & 0.009 & 0.009 & 0.009 & 0.010 & 0.010 & 0.010 & 0.010 \\
\hline\hline
    & 1 & 0.085 & 0.086 & 0.086 & 0.086 & 0.087 & 0.087 & 0.087 & 0.088 & 0.088 \\
    & 2 & 0.030 & 0.030 & 0.031 & 0.031 & 0.031 & 0.031 & 0.032 & 0.032 & 0.032 \\
 10 & 3 & 0.016 & 0.016 & 0.016 & 0.016 & 0.017 & 0.017 & 0.017 & 0.017 & 0.017 \\
    & 4 & 0.010 & 0.010 & 0.010 & 0.010 & 0.010 & 0.010 & 0.010 & 0.011 & 0.011 \\
    & 5 & 0.006 & 0.007 & 0.007 & 0.007 & 0.007 & 0.007 & 0.007 & 0.007 & 0.007 \\
\hline\hline
\end{tabular}
\label{tab:Knp}
\end{table}

\newpage

\end{document}